\documentclass[copyright,creativecommons]{eptcs}
\providecommand{\event}{GANDALF 2010} % Name of the event you are submitting to
\usepackage{breakurl}             % Not needed if you use pdflatex only.

\usepackage{latexsym}
\usepackage{amsmath}
\usepackage{gastex}

\newtheorem{theorem}{Theorem}[section]
       
   \newtheorem{lemma}{Lemma}[section]
    \newtheorem{corollary}{Corollary}[section]
    \newtheorem{definition}{Definition}[section]

\newtheorem{claim}{Claim}[section]

\title{On Modal $\mu$-Calculus over Finite Graphs
 with Bounded Strongly Connected Components}
\author{Giovanna D'Agostino
\institute{Universit\`a degli Studi di Udine\\
DIMI (Dipartimento di Matematica e Informatica)\\
Udine, Italy}
\email{dagostin@dimi.uniud.it}
\and
Giacomo Lenzi
\institute{Universit\`a degli Studi di Salerno\\
DMI (Dipartimento di Matematica e Informatica)\\
Fisciano (SA), Italy}
\email{gilenzi@unisa.it}}

\def\titlerunning{$\mu$-Calculus over graphs with bounded s.c.c.}
\def\authorrunning{G. D'Agostino \& G. Lenzi}
\begin{document}
\maketitle

\begin{abstract}
For every positive integer $k$ we consider the class $SCCk$ of all finite graphs whose strongly connected components have size at most $k$. We show that for every $k$, the Modal $\mu$-Calculus fixpoint hierarchy on $SCCk$ collapses to the level $\Delta_2=\Pi_2\cap\Sigma_2$, but not to $Comp(\Sigma_1,\Pi_1)$ (compositions of formulas of level $\Sigma_1$ and $\Pi_1$). This contrasts with the class of all graphs, where $\Delta_2=Comp(\Sigma_1,\Pi_1)$.
\end{abstract}

\section{Introduction}

The subject of this paper is {\em Modal $\mu$-Calculus}, an extension of Modal Logic with operators for least and greatest fixpoints of monotone functions on sets. This logic, introduced by Kozen in \cite{K83}, is a powerful formalism capable of expressing inductive as well as coinductive concepts and beyond (e.g. safety, liveness, fairness, termination, etc.) and is widely used in the area of verification of computer systems, be them hardware or software, see \cite{MC}.

\medskip

Like Modal Logic, the $\mu$-Calculus can be given a Kripke semantics on graphs. It results that on arbitrary graphs, the more we nest least and greatest fixpoints, the more properties we obtain. In other words, on the class of all graphs, the fixpoint alternation hierarchy $(\Sigma_n,\Pi_n,\Delta_n)$ is infinite, see \cite{A99} and \cite{B96}. Whereas the low levels have a clear ``temporal logic'' meaning ($\Pi_1$ gives safety, $\Sigma_1$ gives liveness, $\Pi_2$ gives fairness), the meaning of the higher levels can be understood in terms of {\em parity games} (a formula with $n$ alternations corresponds to a parity game with $n$ priorities).

\medskip

The fixpoint hierarchy may not be infinite anymore if we restrict the semantics to subclasses of graphs. For instance, over the class of all {\em transitive} graphs (the class known as K4 in Modal Logic), the hierarchy collapses to the class $Comp(\Sigma_1,\Pi_1)$, that is, to  compositions of alternation-free formulas, see \cite{AF} and \cite{DL10}. As another example, it is not difficult to show that on finite {\em trees}, the $\mu$-Calculus collapses to the class $\Delta_1=\Sigma_1\cap\Pi_1$.

\medskip

In this paper we are interested in some classes of finite graphs which generalize finite trees and (up to bisimulation) finite transitive graphs, but are not too far from them. Our classes are characterized by having strongly connected components (s.c.c.) of size bounded by a finite constant.

Note that:
\begin{itemize}
\item every finite tree has all s.c.c. of size one, and
\item every finite transitive graph vertex-colored with  $k$ colors is bisimilar to a graph whose s.c.c. have size $k$.
\end{itemize}

In our opinion, the size of the strongly connected components could be an interesting measure of complexity for finite graphs, analogous, but not equivalent, to the important graph-theoretic notion of {\em tree width}, see \cite{H76}, \cite{RS} and \cite{JRST}. Measures of complexity of finite graphs are gaining importance in the frame of {\em Fixed Parameter Complexity Theory}, where many problems intractable on arbitrary graphs become feasible when some parameter is fixed, see \cite{FG}.

\medskip

The purpose of this paper is to determine to what extent the alternating fixpoint hierarchy collapses on finite graphs with s.c.c. of  bounded  size. First we give a $\Pi_2$ upper bound, which by complementation becomes $\Delta_2=\Sigma_2\cap\Pi_2$. Then we show that the $\Delta_2$ bound is tight, in the sense that already on finite graphs with s.c.c. of size one, the $\mu$-Calculus does not collapse to $Comp(\Sigma_1,\Pi_1)$, that is, to compositions of alternation-free formulas. The latter
can be considered as a level very close to $\Delta_2$ in the alternation hierarchy. In fact, $\Delta_2$ includes $Comp(\Sigma_1,\Pi_1)$ (in arbitrary classes of graphs), and the two levels  coincide on the class of all graphs (see \cite{VK}).

\subsection{Related work}

This paper concerns expressiveness of the $\mu$-Calculus in subclasses of graphs, a subject already treated in previous papers. We mention some of them.

An important theorem in the area (despite it predates the invention of Modal $\mu$-Calculus)
is the De Jongh-Sambin Theorem, see \cite{Smo85}. The theorem considers the important modal logic GL (G\"odel-L\"ob); this logic, besides being deeply studied as a logic of provability, corresponds to a natural class of graphs, i.e. the transitive, wellfounded graphs. The theorem says that fixpoint modal equations in GL have a unique solution. From the theorem it follows that in GL, the $\mu$-Calculus collapses to Modal Logic, see \cite{VB06} and \cite{V05}. For a proof of this collapse independent of the De Jongh-Sambin Theorem,  see \cite{AF09}; in that paper the collapse is also extended to an extension of the $\mu$-Calculus, where fixpoint variables are not necessarily in positive positions in the formulas.

The work \cite{AF} contains a proof of the collapse of the $\mu$-Calculus to the alternation free fragment over transitive graphs (different proofs of this collapse can be found in \cite{DO09} and \cite{DL10}, see  below);  the $\mu$-Calculus hierarchy is also studied in other natural classes of graphs, such as the symmetric and transitive class, where it collapses to Modal Logic, and the reflexive class, where the hierarchy is strict.

Recall that \cite{VB76} characterizes Modal Logic as the bisimulation invariant fragment of First Order Logic, and that likewise, \cite{JW96} characterizes the $\mu$-Calculus as the bisimulation invariant fragment of Monadic Second Order Logic. The work \cite{DO09} extends the results of \cite{VB76} and \cite{JW96} to several subclasses of graphs, including transitive graphs, rooted graphs, finite rooted graphs, finite transitive graphs, well-founded transitive graphs, and finite equivalence graphs (all these classes except the first one are not first order definable, so classical model theory cannot be directly applied; rather, \cite{DO09} uses Ehrenfeucht style games). An unexpected behavior arises over finite transitive frames: the bisimulation invariant fragments of First Order and Monadic Second Order Logic coincide, despite $\mu$-Calculus and Modal Logic do not coincide. These fragments are characterized in \cite{DO09} by means of suitable modal-like operators. From  the above results the authors  obtain the collapse of the $\mu$-Calculus over transitive frames, as well as the inclusion of the $\mu$-calculus in First Order Logic over finite transitive frames.

Finally we mention that \cite{DL10} gives a  proof of the   first order definability of the $\mu$-Calculus over finite transitive frames which is independent   from the work in \cite{DO09}, and contains
a particular case of Theorem \ref{thm:upper} below (namely, the case of the graphs called  ``simple'' in \cite{DL10}, i.e. such that every s.c.c. has at most one vertex for each possible color).

\section{Preliminaries on Modal $\mu$-Calculus}

\subsection{Syntax}

The syntax of a $\mu$-Calculus formula $\phi$ (in negation normal form) is the following:
$$\phi::=\ X\ |\ P\ |\ \neg P\ |\ \phi_1\vee\phi_2\ |\ \phi_1\wedge\phi_2\ |\ \Diamond \phi\ |\ \Box  \phi\ |\  \mu X.\phi\ |\ \nu X.\phi,$$
where $X$ ranges over a countable set $FV$ of fixpoint variables, and $P$ ranges over a countable set $At$ of atomic propositions.

\medskip

The boolean connectives are $\neg$ (negation), $\wedge$ (conjunction) and $\vee$ (disjunction). The modal operators are $\Diamond$ (diamond) and $\Box $ (box).

Finally, there are the fixpoint operators $\mu$ and $\nu$. Intuitively, $\mu X.\phi(X)$ denotes the least fixpoint of the function $\phi$ (a function mapping sets to sets), and $\nu X.\phi(X)$ denotes the greatest such fixpoint.

\medskip

Note that negation is applied only in front of atomic propositions. So, not every formula has a negation. However, every sentence (i.e., every formula without free variables) does have a negation, obtained by applying the De Morgan dualities between the pairs $\wedge$ and $\vee$,   $\Diamond$ and $\Box $, and $\mu$ and $\nu$ (the last duality is given by
$\neg\mu X.\phi(X)=\nu X.\neg\phi(\neg X)$).

\medskip

Free and bound fixpoint variables, as well as scopes of fixpoint operators,  can be defined in complete analogy with First Order Logic (where fixpoint operators are treated in analogy with first order quantifiers).

\medskip

The formulas of the $\mu$-Calculus can be composed in a natural way. Let $\phi$ be a formula and let $P$ be an atom of $\phi$. Suppose that $\psi$ is a formula free for $P$ in $\phi$ (that is, $\psi$ has no free variables $X$ such that some occurrence of $P$ is in the scope of some fixpoint $\mu X$ or $\nu X$). Then we can replace $P$ with $\psi$ everywhere in $\phi$. We obtain a $\mu$-calculus formula $\chi$ which we call the {\em composition} of $\phi$ and $\psi$ (with respect to the atom $P$).

\subsection{Fixpoint hierarchy}

The $\mu$-Calculus formulas can be classified according to the alternation depth of their fixpoints. Formally we have a hierarchy of classes
 $\Sigma_n,\Pi_n,\Delta_n$ as follows.

\medskip

First, $\Sigma_0=\Pi_0$ is the set of the formulas without fixpoints.

\medskip

 Then, $\Pi_{n+1}$ is the smallest class containing $\Sigma_n\cup\Pi_n$ and closed under composition and $\nu$ operators.

 Dually, $\Sigma_{n+1}$ is the smallest class containing $\Sigma_n\cup\Pi_n$ and closed under composition and $\mu$ operators.

\medskip

Note that a property is in $\Pi_n$ if and only if its negation is in $\Sigma_n$, and conversely. In this sense, the classes $\Sigma_n$ and $\Pi_n$ are dual.

\medskip

Finally, a property is said to be in $\Delta_n$ if it is both in $\Sigma_n$ and in $\Pi_n$.

\medskip

The {\em alternation depth} of a $\mu$-Calculus definable property  is the least $n$ such that the property is in $\Sigma_n\cup\Pi_n$.

\subsection{Graphs and trees}

A {\em (directed) graph} is a pair $G=(V,R)$, where $V$ is a set of vertices and $R$ is a binary edge relation on $V$. Sometimes we denote $V$ by $V(G)$ and $R$ by $R(G)$.

Likewise, an {\em undirected graph} is a pair $G=(V,S)$ where $V$ is a set of vertices and $S$ is a symmetric relation on $V$. That is, $xSy$ must imply $ySx$.

Note that to every directed graph we can associate the underlying undirected graph, by letting $S=R\cup R^{-1}$ (i.e. $S$ is the symmetric closure of $R$).

\medskip

A {\em successor} of a vertex  $v$ in $G$ is a vertex $w$ such that $vRw$. The set $Succ(v)$ is the set of all successors of $v$ in $G$. We also say that $v$ is a {\em predecessor} of $w$.

\medskip

A {\em path of length $n$} in a graph $G$ from $v$ to $w$ is a finite  sequence $v_1,v_2,\ldots,v_n$ of vertices such that $v_1=v$, $v_n=w$ and $v_i R v_{i+1}$ for  $1\leq i<n$. A {\em descendant} of $v$ is a vertex $w$ such that there is a path from $v$ to $w$.

\medskip

The {\em strongly connected component} of a vertex $v\in V$ in a graph $G$ is $v$ itself plus the  set of all $w\in V$ such that there is a path from $v$ to $w$ and conversely.

For a positive integer $k$, we denote by $SCCk$ the class of all finite graphs whose strongly connected components have size at most $k$.

\medskip

A {\em tree} is a graph $T$ having a vertex $r$ (the root) such that for every vertex $v$ of $T$ there is a unique path from $r$ to $v$.

The {\em height} of a vertex $v$ of a tree $T$ is the length of the unique path from $r$ to $v$.

A {\em subtree} of a tree $T$ is a subset $U$ of $T$ which is still a tree with respect to the induced edge relation $R(T)\cap U^2$.

\medskip

If $Pred$ is a set of unary predicates, a {\em $Pred$-colored graph} is a graph $G$ equipped with a ``satisfaction'' relation $Rsat\subseteq Pred\times V(G)$, which intuitively specifies which unary predicates are true in which vertices.

One also thinks of the set $Powerset(Pred)$ as a set of ``colors'' of the vertices of $G$, where the color of $v$ is the set of all predicates $P\in Pred$ such that $P\ Rsat\ v$ holds.

\medskip

A {\em pointed graph} is a graph equipped with a distinguished vertex. Similarly one defines pointed colored graphs.

\subsection{Semantics}

Like in usual Modal Logic, the formulas of the $\mu$-Calculus can be interpreted on (colored pointed) graphs via Kripke semantics. One defines inductively a {\em satisfaction} relation between graphs and formulas. The clauses of the satisfaction relation are the usual ones for Modal Logic, plus two new rules which are specific for fixpoints.

\medskip

A pointed, $At\cup FV$-colored graph $(G,Rsat,v)$ satisfies an atom $P$ if $P\ Rsat\ v$ holds, satisfies
$\neg P$ if it does not satisfy $P$, and satisfies a fixpoint variable $X$ if $X\ Rsat\ v$ holds.

\medskip

For the boolean clauses, $(G,Rsat,v)$ satisfies $\phi\wedge	\psi$ if it satisfies $\phi$ and $\psi$; and it satisfies $\phi\vee\psi$ if it satisfies $\phi$ or $\psi$.

\medskip

For the modal clauses, $(G,Rsat,v)$ satisfies $\Diamond\phi$ if there is $w$ with $vRw$ and $(G,Rsat,w)$ satisfies $\phi$; and it satisfies $\Box \phi$ if for every $w$ with $vRw$ we have that $(G,Rsat,w)$ satisfies $\phi$.

\medskip

For the fixpoint clauses, the idea is that $\mu X.\phi(X)$ and $\nu X.\phi(X)$ denote sets which are the least and greatest solutions of the fixpoint equation $X=\phi(X)$, respectively.

\medskip

Formally, $(G,Rsat,v)$ satisfies a
formula $\mu X.\phi$ if $v$ belongs to every set $E$ equal to
 $\phi(E)$, where $\phi(E)$ is the set of all vertices $w$ such that $(G,Rsat[X:=E],w)$ satisfies $\phi$, and where $Rsat[X:=E]$ is the same relation as $Rsat$, except that $X\ Rsat[X:=E]\ z$ holds if and only if $z\in E$.

Dually, $(G,Rsat,v)$ verifies a formula $\nu X.\phi(X)$ if $v$ belongs to some set $E$ equal to $\phi(E)$.

\medskip

A kind of ``global'' modalities are $\Box ^*\phi=\nu X.\phi\wedge \Box X$ and the dual
$\Diamond^*\phi=\mu X.\phi\vee\Diamond X$. The former means that $\phi$ is true ``always'' (i.e. in all descendants of the current vertex), and the latter means that $\phi$ is true ``sometimes'' (i.e. in some descendant).

\subsection{Bisimulation}

Bisimulation between graphs is a generalization of isomorphism, which is intended to capture the fact that two graphs have the same observable behavior.

\medskip

A {\em bisimulation} between two ($Pred$-colored) graphs $G,H$ is a relation $B\subseteq V(G)\times V(H)$, such that if $vBw$ holds, then:
\begin{itemize}
\item $v$ and $w$ satisfy the same predicates in $Pred$;
\item if $vRv'$ in $G$, then there is $w'\in H$ such that $wRw'$ in $H$ and $v'Bw'$;
\item dually, if $wRw'$ in $H$, then there is $v'\in G$ such that $vRv'$ in $G$ and $v'Bw'$.
\end{itemize}

Two pointed, colored graphs $(G,v)$ and $(H,w)$ are called {\em bisimilar} if there is a bisimulation $B$ between $G$ and $H$  such that $vBw$.

\medskip

Every pointed graph $(G,v)$ is bisimilar to a tree, and there is a canonical such tree, called the {\em unfolding} of $(G,v)$, denoted by $U(G,v)$. The vertices of $U(G,v)$ are the finite paths of $G$ starting from $v$. There is an edge from $\pi$ to $\pi'$ if $\pi'$ is obtained from $\pi$ by adding one step at the end. A path $\pi$ satisfies
a predicate  if and only if its last vertex does.
It results that the function mapping a path to its last vertex is a bisimulation between $U(G,v)$ and $(G,v)$.

\medskip

Like Modal Logic, the $\mu$-Calculus is invariant under bisimulation (in fact it can be viewed as a kind of infinitary modal logic).
In particular, every $\mu$-Calculus formula which is valid on all trees is valid on all graphs as well.

\subsection{Tree width}

In this subsection we define tree decompositions and tree width of an undirected graph $G=(V,S)$.

\medskip

Intuitively, the tree width of a graph measures how far the graph is from being a tree.
Being close to a tree is a virtue, because many graph theoretic problems become much easier when restricted to trees.

\medskip

For the benefit of software verification, \cite{O03} argues that programs in many programming languages have control flow diagrams with low tree width (as long as no {\tt goto} command or similar is used).

\medskip

Formally, a {\em tree decomposition} of the graph $G$ is  a pair $({\cal X}, T)$, where ${\cal X} =\{X_1, \ldots, X_n\}$ is a family of subsets of $V$, and $T$ is a tree whose nodes are the subsets $X_i$, satisfying the following properties:

\begin{itemize}
\item The union of all sets $X_i$ equals $V$. That is, each graph vertex is associated with at least one tree node.
\item For every edge $(v, w)$ in the graph, there is a subset $X_i$ that contains both $v$ and $w$. That is, vertices are adjacent in the graph only when the corresponding subtrees have a node in common.
\item If $X_i$ and $X_j$ both contain a vertex $v$, then all nodes $X_z$ of the tree in the (unique) path between $X_i$ and $X_j$ contain $v$ as well. That is, the nodes associated with vertex $v$ form a connected subset of $T$.
\end{itemize}

The {\em width} of a tree decomposition is the size of its largest set $X_i$ minus one. The {\em tree width} $tw(G)$ of a graph $G$ is the minimum width among all possible tree decompositions of $G$.

In this paper, we define the tree width of a {\em directed} graph as the tree width of the
underlying undirected graph.
We denote by $TWk$ the class of all finite directed graphs whose tree width is at most $k$.

\medskip

As a first remark, the tree width of a tree is one (the definition is adjusted so that this is true). In fact, as a tree decomposition we can take all edges of the tree.

Moreover, tree width does not change if we add or remove loops (i.e. edges $(v,v)$) to the graph.

\medskip

Less trivially, we have examples of applications of tree width in the following areas:
\begin{itemize}
\item Robertson-Seymour Graph Minors Theory, see \cite{RS} and \cite{RS86};
\item Complexity Theory, e.g. the Hamiltonian path problem can
be solved in polynomial time if the directed tree width is bounded by a constant, see \cite{JRST}, where the directed tree width is a variant of tree width tailored for directed graphs.
\end{itemize}

\section{Model checking and parity games}

The {\em $\mu$-Calculus model checking problem} is the following algorithmic problem: given a formula $\phi$ of Modal $\mu$-Calculus and a finite graph $G$, decide whether $\phi$ is true in $G$.

A kind of games closely related to the $\mu$-Calculus model checking problem is {\em parity games}. In fact, checking a $\mu$-Calculus formula in a finite graph is a problem computationally equivalent (in polynomial  time) to solving a finite parity game.

\medskip

Parity games can be described as follows. There are two players, let us call them $Odd$ and $Even$. Let $G$ be a countable graph. Let $\Omega:V(G)\rightarrow \omega$ be a priority function with finite range. Let $v_0$ be a starting vertex. The two players move along the edges of the graph. On odd positions, player $Odd$ moves, and  on even positions, player $Even$ moves.

\medskip

If either player has no move, the other wins. Otherwise, the play is an infinite sequence of vertices $v_0,v_1,v_2,v_3\ldots$, and we say that player $Even$ wins the play if the smallest number occurring infinitely often in the sequence $\Omega(v_0),\Omega(v_1),\Omega(v_2),\Omega(v_3)\ldots$ is even. Otherwise, we say that player $Odd$ wins.

\medskip

A {\em strategy} $\cal S$ of a player $Pl$
  is a function from finite
 sequences of vertices $v_0,v_1,v_2,v_3\ldots v_k$, where $v_k$ is a $Pl$-vertex,
 to a successor of $v_k$. A strategy $\cal S$ of $Pl$ is {\em
 winning} if $Pl$ wins all the play which respect $\cal S$.

\medskip

Parity games can be encoded as Borel games in the sense of Descriptive Set Theory; so, by Martin's Borel Determinacy Theorem, see \cite{M75}, parity games are {\em determined}: that is, there is always a player which has a winning strategy in the game.

\medskip

A strategy $\cal S$ of a player is called {\em positional} if
 ${\cal S}(v_1,v_2,v_3\ldots v_k)$ only depends on  the last vertex
 played $v_k$.

\medskip

Parity games are important because they enjoy the following very strong form of determinacy:

\begin{lemma} (positional determinacy, see \cite{EJ})  \label{lemma:det}
If either player has a winning strategy in a parity game, then he has a positional winning strategy.
\end{lemma}

Given that model checking and parity games are polynomial time equivalent, one is solvable in polynomial time if and only if the other is. And given the importance of $\mu$-Calculus for system verification, the polynomial time solvability of these problems is a crucial problem in the area.

\medskip

It is known that the two problems are in the complexity class $UP$ (standing for Unique $P$), that is, the problems solvable in polynomial time by a nondeterministic Turing machine having at most one accepting computation on each input, see \cite{J98}. Note that $UP$ is a subclass of $NP$, and a $co-UP$ bound follows by complementation.

\medskip

Several algorithms have been proposed, starting from the first model checking algorithm of
\cite{EL86}; the working time of this algorithm is $O(m\cdot n^{d+1})$, where $m$ is the size of $\phi$, $n$ is the size of $G$ and $d$ is the alternation depth of $\phi$.

\medskip

Subsequently, \cite{LBC+94} improved the complexity of the Emerson-Lei algorithm to $O(m\cdot n^{\lceil{d/2}\rceil+1})$.

\medskip

Then we have an algorithm
 which works ``fast'' on graphs of bounded tree width (see \cite{O03}). Recall that Courcelle's theory of monadic second order logic \cite{Cou90} implies that on graphs of bounded tree-width $k$, the
model checking problem can be solved in time linear in the size of the graph, that is, the time is $O(n)$. However, the constant hidden in the $O$ (depending on the formula and on the tree width) is  large according to Courcelle's bound. \cite{O03} manages to reduce to time $O(n\cdot (km)^2\cdot d^{2((k+1)m)^2})$, so a little more than exponential in $d,k,m$.

\medskip

For the general case,  the best we have so far is a subexponential algorithm (see \cite{JPZ}), and a general polynomial algorithm is actively searched.

\section{Automata}

\subsection{Parity automata}

Since Rabin automata were introduced in \cite{R69}, tree automata have been studied as
``dynamic'' counterparts of various logics. For instance,
parity automata are expressively equivalent to $\mu$-Calculus formulas, and can be viewed as a ``dynamic'' normal form of the $\mu$-Calculus.

\medskip

There are several equivalent definitions for parity automata, especially differing in the transition function. We choose the following definition.

\medskip

A {\em parity automaton} is a tuple $A=(Q,\Lambda,\delta,q_0,\Omega)$ where:
\begin{itemize}
\item $Q$ is a finite set of states;
\item $\Lambda=Powerset(Pred)$ is the alphabet, where $Pred$ is a finite set;
\item $q_0\in Q$ is the initial state;
\item $\Omega:Q\rightarrow\omega$ is the priority function;
\item $\delta:Q\times\Lambda\rightarrow Dc(Q)$ is the transition function, where $Dc(Q)$ is the set of all disjunctions of  ``cover'' operators
$$cover(q_1,\ldots,q_n)=\Diamond q_1\wedge\ldots\wedge \Diamond q_n\wedge \Box (q_1\vee\ldots\vee q_n),$$
with $q_1,\ldots,q_n\in Q$.\end{itemize}

A semantic game (in fact a kind of parity game) can be defined from an automaton $A$ and a countable, pointed, $Pred$-colored graph $(G,Rsat,v_0)$.

\medskip

Let $V=V(G)$. For $v\in V$, let
$color(v)$ the set of the elements $P\in Pred$ such that $P\ Rsat\ v$. This gives a function $color:V\rightarrow \Lambda$.

\medskip

The players are called $Duplicator$ and $Spoiler$.
Positions of the game are, alternately, elements of $Q\times V$ and subsets of $Q\times V$.

\medskip

The initial position is $(q_0,v_0)$.
On a position $(q,v)$, $Duplicator$ moves by choosing a
  ``marking'' function $m$ from
$Succ(v)$ to $Powerset(Q)$  which, viewed as an interpretation for the
 atoms $Q$ over the graph $\{v\}\cup Succ(v)$,  satisfies the modal
 formula
 $\delta(q,color(v))$.
 $Spoiler$ then moves by choosing  a pair $(q',v')\in m$ with $v'\in Succ(v)$;    the new position becomes $(q',v')$,  and so on.

\medskip

If ever some player has no moves, the other wins. Otherwise, we have an infinite sequence $$(q_0,v_0),m_1,(q_1,v_1),m_2,(q_2,v_2),\ldots,$$ and $Duplicator$ wins if in the sequence
$\Omega(q_0),\Omega(q_1),\Omega(q_2),\ldots$, the least integer occurring infinitely often is even. Otherwise, $Spoiler$ is the winner.

\medskip

The automaton $A$ {\em accepts} the graph $G$ if $Duplicator$ has a winning strategy in the game of $A$ on $G$. The {\em language} defined by $A$ is the set of all graphs accepted by $A$.

\medskip

If $q$ is a state of the automaton $A$, we denote by $(A,q)$ the automaton like $A$ except that the initial state is $q$.

\medskip

Like in every two player game, if $\cal S$ is a strategy of either player in an automaton game, the moves of $\cal S$ can be organized in a tree, called the {\em strategy tree} of $\cal S$.

In particular,    if $\cal S$ is a strategy for $Duplicator$  on a graph $G$, the  strategy tree  of $\cal S$   can be represented as a labeled tree  as follows. The  nodes are all possible  finite prefixes $(q_0,v_0)m_1(q_1,v_1) m_2 \ldots (q_n,v_n)$ of a play (ending in a move of $Spoiler$)  where $Duplicator$ uses $\cal S$, with the node
$(q_0,v_0)m_1(q_1,v_1) m_2 \ldots (q_n,v_n)$ being a successor of the node
$(q_0,v_0)m_1(q_1,v_1) m_2 \ldots (q_{n-1},v_{n-1})$. The {\em label} of the node $$(q_0,v_0)m_1(q_1,v_1) m_2 \ldots (q_n,v_n)$$  is the pair $(q_n,v_n)$.

\medskip

Since the transition function are disjunctions of covers, it follows that  if $T$ is a  strategy tree for $Duplicator$ on a graph $G$, then the second (vertex) components of the labels of the nodes of $T$  form   a tree bisimilar to $G$.

\medskip

In the following, and in particular in Section  \ref{sect:upper}, we shall need more general automata, where,  besides covers, among the disjuncts of the  transition function $\delta(q,c)$ we may also find   conjunctions of diamonds:
$$
\Diamond(q_1)\wedge\dots \wedge \Diamond(q_n).
$$
This kind of automata, however, can be simulated  by  ``cover-automata'', in the following way.

Suppose $A$ is such an automaton.
\begin{itemize}
\item First, add a new state $q_t$   with $\Omega(q_t)=0$ and
$\delta(q_t,c)=cover(q_t)\lor cover(\emptyset)$ (notice that, starting from  $q_t$, the new automaton accepts any graph).
\item
Then,
substitute any disjunct  having the form  $\diamond q_1\wedge \ldots \wedge \diamond q_n$    with $cover(q_1,\ldots,q_n,q_t)$.
\end{itemize}

The new automaton   only uses  disjuctions of ``covers'' in the transition function, and is equivalent to $A$.

\medskip

Notice finally that the game of a parity automaton on a graph can be coded into a parity game, hence parity automata enjoy  positional determinacy by Lemma \ref{lemma:det}. This is a good reason to choose parity automata rather than other, expressively equivalent kinds of automata.

\subsection{Weak parity automata}

A parity automaton is called {\em weak} if for every $(q,\lambda)\in Q\times\Lambda$ and every state $q'$ occurring in $\delta(q,\lambda)$, one has $\Omega(q')\leq\Omega(q)$. So,
along every transition, the priority does not increase. This implies that in every infinite play, the priority is eventually constant, and $Duplicator$ wins if and only if this eventual priority is even.

\medskip

Weak parity automata are expressively equivalent, on arbitrary graphs, to compositions of $\Sigma_1$ and $\Pi_1$ formulas of the $\mu$-Calculus.

\subsection{B\"uchi automata}

A {\em B\"uchi automaton} is a parity automaton where $\Omega:Q\rightarrow\{0,1\}$. When talking about B\"uchi automata, one calls {\em final} a state $q$ such that $\Omega(q)=0$.
Then $Duplicator$ wins an infinite play if and only if the play visits final states infinitely often.

\medskip

Note that a B\"uchi automaton with conjunctions of diamonds is equivalent to a cover B\"uchi automaton, because adding a state $q_t$ with $\Omega(q_t)=0$ to a B\"uchi automaton produces an automaton of the same class.

\medskip

In the $\mu$-Calculus fixpoint hierarchy, B\"uchi automata coincide with the class $\Pi_2$.

\subsection{coB\"uchi automata}

The dual of B\"uchi automata are co-B\"uchi automata.

\medskip

A {\em coB\"uchi automaton} is a parity automaton where $\Omega:Q\rightarrow\{1,2\}$. When talking about coB\"uchi automata, one calls {\em final} a state $q$ such that $\Omega(q)=2$.
Then $Duplicator$ wins an infinite play if and only if the play visits final states always except for a finite number of times.

\medskip

In the $\mu$-Calculus fixpoint hierarchy, coB\"uchi automata coincide with the class $\Sigma_2$.

\section{The upper bound}\label{sect:upper}

\begin{theorem}\label{thm:upper} For every $k$, every B\"uchi automaton is equivalent in $SCCk$ to a coB\"uchi automaton.
\end{theorem}

 {\em Proof:} let $B$ be a B\"uchi automaton. Let $Q$ be the set of states of $B$. By Lemma \ref{lemma:det}, if $Duplicator$ has a winning strategy for $B$ in a graph $G$ of class $SCCk$, then he or she has a positional winning strategy, call it ${\cal S}_p$.

\medskip

Let $\pi$ be an infinite play of ${\cal S}_p$. Then $\pi$ must have, from a certain point on, at least a final state every $|Q|k$ moves. In fact, if this were not true, then $\pi$ would have infinitely many nonfinal subsequences of size $|Q|k+1$. Since $G$ is finite, $\pi$ eventually enters some s.c.c. $S$ where it remains forever. If we take $|Q|k+1$ consecutive nonfinal moves in $S$, then since $S$ has at most $k$ elements, by the pigeonhole principle there is a repeated pair $(q,v),\ldots,(q,v)$ among these moves. Now if $Spoiler$ repeats the moves he or she did between the two equal pairs above, $Duplicator$ is also forced (in ${\cal S}_p$) to repeat his or her moves, because ${\cal S}_p$ is positional. So, ${\cal S}_p$ has an infinite play with only finitely many nonfinal states, contrary to the fact that ${\cal S}_p$ is winning for $Duplicator$ in the B\"uchi automaton $B$.

\medskip

Summing up, if $Duplicator$ manages to have infinitely many final states in a play, then he or she manages to have final states at most every $|Q|k$ moves, form a certain moment on. This corresponds to the coB\"uchi automaton $C$ which we are going to define.

\medskip

The idea is to play $B$ and to memorize the last $|Q|k$ states of the play.

\medskip

The alphabet of $C$ will be the same of $B$.

\medskip

The states of $C$ will be the nonempty lists of states of $B$ with length at most $|Q|k$.

\medskip

The initial state of $C$ is the list of length one  consisting of the initial state of $B$.

\medskip

The final states of $C$ will be the lists of length $|Q|k$ containing at least one final state of $B$.

\medskip

Finally, the transition function $\delta_C$ of $C$ will mimic the function $\delta_B$ of $B$ while memorizing the last $|Q|k$ states. Formally, we say that a marking $m$ satisfies $\delta_C(L,\gamma)$ if verifies a formula of the kind $$cover(L'q_1,\ldots,L'q_n),$$ where:

\begin{itemize}
\item $cover(q_1,\ldots,q_n)$ is a disjunct of
$\delta_B(last(L),\gamma)$, and
\item $L'=L$ if $L$ has length less than $|Q|k$, and $L'$ is $L$ minus the first element  otherwise.
\end{itemize}

Now if $B$ accepts a graph $G$ then, as we have seen, there is a winning strategy ${\cal S}_p$ of $Duplicator$ where finals repeat every $|Q|k$ times from a certain point on. So,
$C$ also accepts $G$, with the strategy consisting of playing ${\cal S}_p$, and memorizing the last $|Q|k$ states of the play.

\medskip

Conversely, if $C$ accepts a graph $G$, via any winning strategy ${\cal S}'$ of $Duplicator$, then in ${\cal S}'$, final states of $B$ occur infinitely often in every infinite play,
so $B$ also accepts $G$ with the strategy consisting of taking the last components of the lists of ${\cal S}'$.

\medskip

So, the automata $B$ and $C$ are equivalent.

\rightline{Q.E.D.}

\begin{corollary} For every $k\geq 1$, the $\mu$-Calculus collapses in $SCCk$ to
$\Delta_2=\Sigma_2\cap\Pi_2$.
\end{corollary}

{\em Proof:} we show by induction on $n\geq 2$ that $\Sigma_n$ and $\Pi_n$ collapse to $\Delta_2$. For $n=2$, $\Sigma_2$ is included in $\Pi_2$, so $\Sigma_2$ is included in $\Delta_2$. $\Pi_2$ is analogous.

\medskip

For $n\geq 2$, consider $\Sigma_{n+1}$. This class is the closure of $\Sigma_n\cup\Pi_n$ with respect to composition and $\mu$; by inductive hypothesis, $\Sigma_n\cup\Pi_n$ coincide with $\Sigma_2$, so $\Sigma_{n+1}$ is the closure of $\Sigma_2$ with respect to composition and $\mu$, that is, $\Sigma_{n+1}$ coincides with $\Sigma_2$, hence it collapses to $\Delta_2$.

\medskip

Likewise, consider $\Pi_{n+1}$. This class is the closure of $\Sigma_n\cup\Pi_n$ with respect to composition and $\nu$; by inductive hypothesis, $\Sigma_n\cup\Pi_n$ coincide with $\Pi_2$, so $\Pi_{n+1}$ is the closure of $\Pi_2$ with respect to composition and $\nu$, that is, $\Sigma_{n+1}$ coincides with $\Pi_2$, hence it collapses to $\Delta_2$.

\rightline{Q.E.D.}

\section{The lower bound}\label{sect:lower}

\begin{theorem}\label{thm:lower}
There is a B\"uchi automaton which is not equivalent in $SCC1$ to any weak parity automaton.
\end{theorem}

{\em Proof:} the proof needs some definitions and lemmas.

\begin{definition}

Let $F$ be a predicate (standing for final).
Let $(G,v_0)$ be a pointed, $F$-colored graph. This means that each vertex can satisfy $F$ (in which case we call it an {\em $F$-vertex}) or not (in which case we call it {\em $N$-vertex}, $N$ standing for nonfinal).

 We define the following (parity-like) game $\Gamma(G,v_0)$ on $G$. Call $PN$ and $PF$ two players. The positions are the vertices of $G$. The initial position is $v_0$. On $N$ vertices, player $PN$ moves along one edge. On $F$ vertices, likewise, player $PF$ moves along one edge.

 If either player has no move, the other wins. Otherwise, the play is infinite, and player $PN$ wins if the play visits $F$ vertices infinitely often, and player $PF$ wins otherwise (this interchange between players $PN$ and $PF$ in the definition of the winning condition seems to be necessary for the argument to work).

\end{definition}

For convenience, let us say that a graph $(G,v_0)$ {\em verifies property $\Gamma$} if and only if player $PN$ has a winning strategy in the game $\Gamma(G,v_0)$.

\medskip

\begin{lemma}
The property $\Gamma$ is B\"uchi-expressible.
\end{lemma}

{\em Proof:} consider the following B\"uchi automaton $B_{\Gamma}$.

\medskip

The only predicate is  $F$, whose negation we denote by $N$.

\medskip

There are two states $q_N$ and $q_F$ plus an initial state $q_0$.

\medskip

 We decree that $q_F$ is final and $q_N$ is nonfinal (the priority of $q_0$ is irrelevant, let us decide that $q_0$ is final).

\medskip

Finally, the transition function $\delta_{\Gamma}$ of $B_{\Gamma}$ is the following:

\begin{itemize}
\item $\delta_{\Gamma}(q_0,N)=\delta_{\Gamma}(q_N,N)=(\Diamond q_N)\vee
(\Diamond q_F)$;
\item $\delta_{\Gamma}(q_0,F)=\delta_{\Gamma}(q_F,F)=\Box(q_N\vee q_F)$;
\item $\delta_{\Gamma}(q_F,N)=\delta_{\Gamma}(q_N,F)=false$ (the empty disjunction).
\end{itemize}

Note that $B_{\Gamma}$ is equivalent to $\Gamma$.

\medskip

In fact, every winning strategy $\cal S$ for $Duplicator$ in the automaton $B_{\Gamma}$ in a graph $G$ can be translated into a winning strategy ${\cal S}'$ for player $PN$ in $\Gamma(G)$, which consists in choosing any successor of the current vertex which is marked $q_N$ or $q_F$ in $\cal S$ (assuming that this current vertex is an $N$ vertex).

\medskip

Conversely, we   translate  a strategy  ${\cal S}'$ winning for player $PN$ in $\Gamma(G)$ into  a strategy $\cal S$ winning for $Duplicator$ in $B_{\Gamma}$,   as follows.

In a position $(q_0,v)$    or $(q_N,v)$ , where $v$ is an  $N$- vertex,  $Duplicator$   takes the vertex $v'$ chosen by ${\cal S}'$ and marks it with $q_N$, if $v'$ is a $N$ vertex, and with $q_F$, if it is an $F$ vertex.

In a position  $(q_0,v)$    or $(q_F,v)$ , where $v$ is an  $F$- vertex,  $Duplicator$ marks all successors of $v$:    with $q_N$, if  the successor  is a $N$ vertex, and with $q_F$, if it is an $F$ vertex (notice that,  following the strategy ${\cal S}'$,  a play will never reach a position of type $(q_N,v)$ for an $F$-vertex, or $(q_F,v)$, for an $N$-vertex).

\rightline{Q.E.D}

\begin{corollary}
 The property $\Box ^*\Gamma$ is B\"uchi expressible.
\end{corollary}

{\em Proof:} $\Box ^*\Gamma$ is the composition of the B\"uchi (hence $\Pi_2$) property $\Gamma$ and of the $\Pi_1$ (hence $\Pi_2$)  formula $\Box ^*P$, where $P$ is an atomic proposition. Since $\Pi_2$ is stable under composition,  the property in question is $\Pi_2$, or equivalently, is B\"uchi expressible.

\rightline{Q.E.D.}

\medskip

Now we show that there is no weak parity automaton equivalent to $\Box ^*\Gamma$ in $SCC1$.

\begin{definition} Let $W$ be an automaton. A state $q$ of $W$ is {\em $\Box ^*\Gamma$-winning} if for every graph $G$ belonging to $SCC1$, if $(W,q)$ accepts $G$, then $G$ verifies $\Box ^*\Gamma$.\\
A graph $G$ {\em witnesses against} $q$ if $G$ belongs to $SCC1$, $(W,q)$ accepts $G$ but $G$ does not verify $\Box ^*\Gamma$ (so $q$ is $\Box ^*\Gamma$-winning if and only if there are no witnesses against $q$).

\end{definition}

\begin{definition}
A {\em   finite   pseudotree} is    a finite graph  obtained from a finite tree by adding loops to some nodes.
\end{definition}

\begin{lemma}\label{lemma:pseudo} Every graph $G$ belonging to $SCC1$ is bisimilar to a finite pseudotree.
\end{lemma}

{\em Proof:} let $G_1$ the graph $G$ where the loops have been removed. Let $G_2$ be the unfolding of $G$, which is a finite tree. Let $H$ be the graph resulting from $G_2$ by attaching a loop to any bisimilar copy of a vertex of $G$ having a loop. $H$ is bisimilar to $G$ and is a finite pseudotree.

\rightline{Q.E.D.}

\begin{lemma} \label{lemma:win}
Suppose $W$ is a (weak) automaton such that the initial state $q_0$ is $\Box^*\Gamma$-winning.

If   $G$ is a finite pseudotree, and $T$ is  a winning strategy tree of $Duplicator$ for $W$ on   $G$,  then all  the states $q$ belonging to a label of $T$ are $\Box^* \Gamma$-winning.
 \end{lemma}

{\em Proof:}
Suppose by way of a contradiction that $T$ is  a winning strategy tree of $Duplicator$ for $W$ on   $G$, but  there exists a node $t\in T$ with label $(q,v)$, such that  $q$ is not $\Box^* \Gamma$-winning.

This means that there exists  a graph $G_q$ in $SCC1$ which is accepted by $(W,q)$ such that $\Box^* \Gamma$ is false in $G_q$. Let $T_q$ be a winning strategy tree for $(W,q)$ on $G_q$.

Consider the tree $T'$ which is  obtained from $T$ by substituting the subtree rooted in  $t$ with $T_q$.

\begin{claim} $T'$ is a strategy tree for $(W,q_0)$ on a finite $SCC1$-graph $G'$  containing a reachable   node $g$ such that $(G,g)$ is isomorphic  to $G_q$.
\end{claim}

{\em Proof:} letting $m$ be the height of $t$ in $T$, do the following:
\begin{itemize}
\item
Replace the subree of $T$ rooted in $t$ with $G_q$, and
\item  for any  node   $s\neq t$ of $T$ with height $m$,    consider its label  $(q_s,v_s)$ and replace the subtree of $T$ rooted in $s$ with   the  subgraph of $G$ consisting of the descendants of  $v_s$.
\end{itemize}

  The resulting graph $G'$ is a graph belonging to $SCC1$ containing a reachable   node $g$ such that $(G,g)$ is isomorphic  to $G_q$.

  Moreover,
$T'$ is a winning strategy tree for $(W, q_0)$ on $G'$.  This proves the claim.

\rightline{Q.E.D.}

\medskip

From the claim,  we get a contradiction:  by hypothesis,  $(W,q_0)$ is equivalent to $ \Box^*\Gamma$, and by the claim, $(W,q_0)$ accepts $G'$; on the other hand,   $\Box^* \Gamma$ is false in $G'$. This proves the lemma.

\rightline{Q.E.D.}

\begin{definition}
 For all natural numbers $k$,  let   $n= {2^k} +1$,  and let $G_k$   be the graph having as set of nodes the set
$$\{  v_i, v_{i,1}, v_{i,2},\ldots  v_{i,n} :0\leq  i<k\} \cup \{ v_k \},$$
 where,   for  $0\leq  i \leq k$,   the nodes  $v_i$  are reflexive $F$ nodes, while the  $v_{i,j}$'s   are     irreflexive nodes satisfying $N$.

Moreover, if  $i<k$    the graph $G_k$  has arches  $(v_i,v_{i,1}), \ (v_{i,1},v_{i,2}),\ldots,(v_{i,n-1},v_{i,n}),( v_{i,n}, v_{i+1})$.

The root is   $v_0$.
\end{definition}

 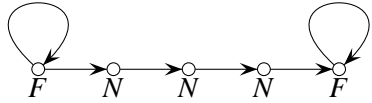
\begin{figure}[ht]
 \begin{center}
\begin{picture}(34,65)(-10,-10)
\gasset{Nw=1.7,Nh=1.7,NLdist=2.5,AHLength=2.5}

% Nw, Nh   --- node width, height
% NLdist --- la distanza tra il nodo ed il label
% NLangle --- dove viene messa la label (0..360 gradi)

\node[NLangle=270,Nmr=1](a)(30,10){$F$}
\node[NLangle=270,Nmr=1](b)(20,10){$N$}
\node[NLangle=270,Nmr=1](c)(10,10){$N$}
\node[NLangle=270,Nmr=1](d)(0,10){$N$}
\node[NLangle=270,Nmr=1](e)(-10,10){$F$}
\drawloop[loopangle=90](a){}
 \drawloop[loopangle=90](e){}
\drawedge(b,a){}
\drawedge(c,b){}
\drawedge(d,c){}
\drawedge(e,d){}
\end{picture}
 \end{center}
\caption{ The graph $G_1$}
 \end{figure}

\medskip

Note that all graphs $G_k$ are pseudotrees and satisfy $\Box^*\Gamma$.

Before passing to the next lemma, let us define $N_{loop}$ to be the graph consisting of one reflexive  $N$ node.

\begin{lemma}\label{lemma:ind}
 For all  $h$  and  $k$ with $h\leq k$, there exists no weak automaton   $W$    with $h$ states having    a positional winning strategy   tree  $T$ of $Duplicator$ on   $G_k$,   where the initial state $q_0$ is $\Box^*\Gamma$-winning.
\end{lemma}

{\em Proof:}  By induction on $k$.

\medskip

 Let $k=1$. Then $W$  has  only one state $q_0$, and there exists      a winning strategy   tree  $T$  for $Duplicator$  on   $G_1$ decorated only by $q_0$   where  $q_0$ is $\Box^*\Gamma$ -winning,  then $\Omega(q_0)$ is even  and $Cover(q_0)$   should be a disjunct of    $\delta(q_0,N)$.

    But then $W$  would accept $N_{loop}$, which does not verify $\Box^* \Gamma$, and $q_0$ would not be $\Box^* \Gamma$-winning, contrary to the hypotheses.

\medskip

Let $k>1$.  Suppose there are  $W$ and $T$ such that $W$ has $h$ states with $h\leq k$ and $T$  is
a winning strategy   tree    for $Duplicator$ in  the $W$-game      on   $G_k$,  where  the initial state $q_0$ is $\Box^* \Gamma$-winning.
By Lemma \ref{lemma:win}, we know that all states appearing as labels in $T$ are $\Box^*\Gamma$-winning.

\begin{claim} There is a node $t\in T$, labeled by $(q,v_0)$ or $(q,v_{0,i})$ for some $i$,  with $\Omega(q)<\Omega(q_0)$.
\end{claim}

{\em Proof:}  Suppose,  by way of a contradiction, that
  all labels  $(q,v_0)$ or $(q,v_{0,i})$ in $T$  have the same priority  $\Omega(q)=\Omega(q_0)$.  First, $\Omega(q_0)$ is even because
\begin{itemize}
\item
  there exists an infinite path in $T$  corresponding to the same node $v_0$, labeled  with states  having the same priority of   $ q_0 $, and
  \item $T$ is winning for $Duplicator$ in $W$, which is a weak automaton.
 \end{itemize}

    \begin{figure}[ht]
\begin{center}
\begin{picture}(34,65)(-10,-10)
%\gasset{Nw=1.7,Nh=1.7,AHLength=2.5}

% Nw, Nh   --- node width, height
% NLdist --- la distanza tra il nodo ed il label
% NLangle --- dove viene messa la label (0..360 gradi)

\node[Nw=1.7,Nh=1.7,AHLength=2.5,NLangle=270,Nmr=1,NLdist=2.5](a)(30,0){$(q_0,v_0)$}
\node[Nw=1.7,Nh=1.7,AHLength=2.5,NLangle=0,Nmr=1,NLdist=10](b)(30,12){$(-,v_{0,1})$}
\node[Nw=1.7,Nh=1.7,AHLength=2.5, NLangle=0,Nmr=1,NLdist=10](c)(30,24){$(-,v_{0,2})$}
 \node[Nw=0.3,Nh=0.3,AHLength=0.3, NLangle=0,Nmr=1,NLdist=10](d)(30,28){}
  \node[Nw=0.3,Nh=0.3,AHLength=0.3, NLangle=0,Nmr=1,NLdist=10](e)(30,30){}
   \node[Nw=0.3,Nh=0.3,AHLength=0.3, NLangle=0,Nmr=1,NLdist=10](f)(30,32){}
     \node[Nw=1.7,Nh=1.7,AHLength=2.5,NLangle=0,Nmr=1,NLdist=10]((g)(30,36){$(-,v_1)$}
      \node[Nw=0.3,Nh=0.3,AHLength=0.3, NLangle=0,Nmr=1,NLdist=10](h)(30,40){}
  \node[Nw=0.3,Nh=0.3,AHLength=0.3, NLangle=0,Nmr=1,NLdist=10](k)(30,42){}
   \node[Nw=0.3,Nh=0.3,AHLength=0.3, NLangle=0,Nmr=1,NLdist=10](l)(30,44){}
    \drawedge(a,b){}
\drawedge(b,c){}

   \node[Nw=1.7,Nh=1.7,AHLength=2.5,NLangle=270,Nmr=1,NLdist=2.5](a')(10,0){$(-,v_0)$}
\node[Nw=1.7,Nh=1.7,AHLength=2.5,NLangle=0,Nmr=1,NLdist=10](b')(10,12){$(-,v_{0,1})$}
\node[Nw=1.7,Nh=1.7,AHLength=2.5, NLangle=0,Nmr=1,NLdist=10](c')(10,24){$(-,v_{0,2})$}
 \node[Nw=0.3,Nh=0.3,AHLength=0.3, NLangle=0,Nmr=1,NLdist=10](d')(10,28){}
  \node[Nw=0.3,Nh=0.3,AHLength=0.3, NLangle=0,Nmr=1,NLdist=10](e')(10,30){}
   \node[Nw=0.3,Nh=0.3,AHLength=0.3, NLangle=0,Nmr=1,NLdist=10](f')(10,32){}
     \node[Nw=1.7,Nh=1.7,AHLength=2.5,NLangle=0,Nmr=1,NLdist=10]((g')(10,36){$(-,v_1)$}
      \node[Nw=0.3,Nh=0.3,AHLength=0.3, NLangle=0,Nmr=1,NLdist=10](h')(10,40){}
  \node[Nw=0.3,Nh=0.3,AHLength=0.3, NLangle=0,Nmr=1,NLdist=10](k')(10,42){}
   \node[Nw=0.3,Nh=0.3,AHLength=0.3, NLangle=0,Nmr=1,NLdist=10](l')(10,44){}

    \drawedge(a',b'){}
\drawedge(b',c'){}

   \node[Nw=1.7,Nh=1.7,AHLength=2.5,NLangle=270,Nmr=1,NLdist=2.5](a")(-10,0){$(-,v_0)$}
\node[Nw=1.7,Nh=1.7,AHLength=2.5,NLangle=0,Nmr=1,NLdist=10](b")( -10,12){$(-,v_{0,1})$}
\node[Nw=1.7,Nh=1.7,AHLength=2.5, NLangle=0,Nmr=1,NLdist=10](c")( -10,24){$(-,v_{0,2})$}
 \node[Nw=0.3,Nh=0.3,AHLength=0.3, NLangle=0,Nmr=1,NLdist=10](d")( -10,28){}
  \node[Nw=0.3,Nh=0.3,AHLength=0.3, NLangle=0,Nmr=1,NLdist=10](e")( -10,30){}
   \node[Nw=0.3,Nh=0.3,AHLength=0.3, NLangle=0,Nmr=1,NLdist=10](f")( -10,32){}
     \node[Nw=1.7,Nh=1.7,AHLength=2.5,NLangle=0,Nmr=1,NLdist=10]((g")( -10,36){$(-,v_1)$}
      \node[Nw=0.3,Nh=0.3,AHLength=0.3, NLangle=0,Nmr=1,NLdist=10](h")( -10,40){}
  \node[Nw=0.3,Nh=0.3,AHLength=0.3, NLangle=0,Nmr=1,NLdist=10](k")( -10,42){}
   \node[Nw=0.3,Nh=0.3,AHLength=0.3, NLangle=0,Nmr=1,NLdist=10](l")( -10,44){}

    \drawedge(a",b"){}
\drawedge(b",c"){}

   \drawedge(a,a'){}
 \drawedge(a',a"){}

  \node[Nw=0.3,Nh=0.3,AHLength=0.3, NLangle=0,Nmr=1,NLdist=10](d")( -4,2){}
  \node[Nw=0.3,Nh=0.3,AHLength=0.3, NLangle=0,Nmr=1,NLdist=10](e")( -6,2){}
   \node[Nw=0.3,Nh=0.3,AHLength=0.3, NLangle=0,Nmr=1,NLdist=10](f")( -8,2){}

    \node[Nw=0.3,Nh=0.3,AHLength=0.3, NLangle=0,Nmr=1,NLdist=10](d")( -4,12){}
  \node[Nw=0.3,Nh=0.3,AHLength=0.3, NLangle=0,Nmr=1,NLdist=10](e")( -6,12){}
   \node[Nw=0.3,Nh=0.3,AHLength=0.3, NLangle=0,Nmr=1,NLdist=10](f")( -8,12){}

    \node[Nw=0.3,Nh=0.3,AHLength=0.3, NLangle=0,Nmr=1,NLdist=10](d")( -4,24){}
  \node[Nw=0.3,Nh=0.3,AHLength=0.3, NLangle=0,Nmr=1,NLdist=10](e")( -6,24){}
   \node[Nw=0.3,Nh=0.3,AHLength=0.3, NLangle=0,Nmr=1,NLdist=10](f")( -8,24){}

    \node[Nw=0.3,Nh=0.3,AHLength=0.3, NLangle=0,Nmr=1,NLdist=10](d")( -4,35){}
  \node[Nw=0.3,Nh=0.3,AHLength=0.3, NLangle=0,Nmr=1,NLdist=10](e")( -6,35){}
   \node[Nw=0.3,Nh=0.3,AHLength=0.3, NLangle=0,Nmr=1,NLdist=10](f")( -8,35){}

\end{picture}
\end{center}
\caption{We show  the tree $T$,  but    for any edge $(v,v')$ in $G_k$ and $t$ in $T$ labeled $(-,v)$ we only draw one successor of $t$ labeled $(-,v')$, whereas there could  be many of them.}
\end{figure}
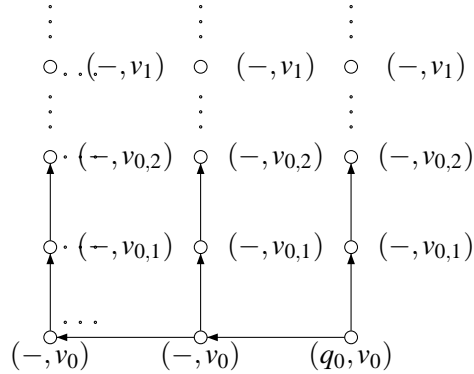

 For $i\leq n$ (where $n= {2^k}+1$), let
$$
Q_i=\{ q\in Q  :   \hbox{  $\exists t\in T$   labeled by  $(q, v_{0,i}) $  } \},
$$
where $Q$ is the set of states of $W$.

Since     all   $Q_i$ are nonempty subsets of $Q$ which has $h$ elements,  and since $2^h<n$, by the pigeonhole principle  there must be two levels  $i<i+j\leq n$ with $Q_i=Q_{i+j}$.
 Fix  $q^*\in Q_i$: we next prove that  $(W,q^*)$   accepts $N_{loop}$ by
 constructing  a winning strategy tree $T_{q^*}^\infty$   for $(W,q^*)$ on $N_{loop}$ as follows.

 For any $q\in Q_i$, consider a
  node $t\in T$  labelled by $(q,v_{0,i})$ and    the subtree $T_{q}$  of $T$ rooted   in $t$ (since we suppose that the strategy for $Duplicator$ is positional, this tree does not depend on $t$, but only on $q$).
  Erase from $T_{q}$  all nodes of height greater than $j$. In this way the leaves of the remaining tree are labeled by pairs $(q',v_{0,i+j})$, for some $q'\in Q_{i+j}=Q_i$.
    Change the second component of the  labels of all nodes of  the resulting tree  to
$ v_{0,i} $, and call $T_{q}^{<j}$  the resulting labeled  finite tree.

  Now consider the fixed state $q^*\in Q_i$.
We define inductively a sequence $T_{q^*}^{m},m=1,2,3,\ldots$ of finite trees.

\begin{itemize}
\item Initially let $T_{q^*}^{1}=T_{q^*}^{<j}$;
\item inductively, given $T_{q^*}^{m}$,
 for all $q\in Q_{i+j}=Q_i$,    attach  to all  leaves of  $T_{q^*}^{m}$  labelled by $(q,v_{0,i})$,     a copy of $T_{q}^{<j}$; call $T_{q^*}^{m+1}$ the result.
\end{itemize}

The sequence of finite trees $T_{q^*}^m$ converges, in the usual sense, to an infinite tree. Let $T_{q^*}^\infty$ be the limit.
 Notice that $T_{q^*}^\infty$ is a winning strategy tree of $Duplicator$ for the automaton $(W,q^*)$ on the graph $N_{loop}$.

Having proved that $(W,q^*)$ accepts $N_{loop}$, we get a contradiction: $N_{loop}$ is a witness against $q^*$, hence $q^*$ should not be $\Box^*\Gamma$ winning, in contradiction with Lemma \ref{lemma:win}. This proves the claim.

\rightline{Q.E.D.}

\medskip

 Using the claim above, we see that there is   a node $t$ in $T$ labeled by $(q_1,v_1)$  for some  $\Box^*\Gamma$-winning state $q_1$  with
$\Omega(q_1)<\Omega(q_0)$ (just follow a path from a node labeled    $(q,v_0)$ or $(q,v_{0,i})$ with $\Omega(q)<\Omega(q_0)$  to a node labeled $(q_1,v_1)$).

Consider now the automaton $W'$ which is $W$ restricted to states which are ``reachable'' using the $W$-transition function  from
$q_1$.   We have:
\begin{itemize}
\item if $h'$ is the number of states of $W'$, then $h'<h$, since $\Omega( q_0)> \Omega(q_1)$ and $W$ is a weak automaton;
\item    the subtree $T'$ of $T$ rooted at $t$    is a winning strategy tree for $W'$ on the graph $G_{k-1}$, and from $h'<h\leq k$ it follows $h'\leq k-1$;
\item $(W',q_1)$ is equivalent to $(W,q_1)$, hence, $q_1$ is a $\Box^*\Gamma$-winning state for $W'$ as well.
\end{itemize}

From  the points above,  we easily obtain a  contradiction by induction.
(As a final  remark, notice  that the positionality hypothesis is not necessary, but has been added to simplify the inductive step). This proves the lemma.

\rightline{Q.E.D.}

\medskip

Now the proof of Theorem \ref{thm:lower} is concluded as follows.

Suppose for an absurdity that there is a weak automaton $W$ equivalent to $\Box ^*\Gamma$ in $SCC1$. Let $k$ be the number of states of  $W$. Consider the graph $G_k$. Then $W$ accepts $G_k$, so there is a winning strategy tree $T$ of $Duplicator$ for $W$ on $G_k$, and the initial state of $T$ is $\Box ^*\Gamma$-winning. But this is in contrast with Lemma \ref{lemma:ind}. So $W$ cannot exist.
\nopagebreak

\rightline{Q.E.D.}

\begin{corollary} \label{cor:coll} Over the class $SCC1$
(hence also over $SCCk$ for any $k>1$) we have:
 \begin{itemize}
  \item $\Delta_2\neq Comp(\Sigma_1,\Pi_1)$;
 \item  the   $\mu$-Calculus does not collapse  to $Comp(\Sigma_1,\Pi_1)$.
 \end{itemize}
 \end{corollary}

\section{Conclusions and future work}

Let us mention a couple of applications of our results. The first application (of Section \ref{sect:upper}) is to the model checking problem:

\begin{corollary} For every $k$, the $\mu$-Calculus model checking problem for a fixed formula $\phi$ is quadratic (i.e. $O(n^2)$) for graphs of class $SCCk$.
 \end{corollary}

{\em Proof:} the algorithm consists in first translating $\phi$ into a B\"uchi automaton (which takes a time depending on $\phi$ and $k$ but not on the graph), and then applying the algorithm of \cite{LBC+94} with $d=2$.

\rightline{Q.E.D.}

\medskip

As a second application (of Section \ref{sect:lower} this time) let us consider tree width:

 \begin{corollary} On the class $TW1$,
the $\mu$-Calculus does not collapse to
 $Comp(\Sigma_1,\Pi_1)$.
 \end{corollary}

 {\em Proof:} Suppose for an absurdity that
 $\mu=Comp(\Sigma_1,\Pi_1)$    on $TW1$. Now every pseudotree has an underlying undirected graph  of tree width one. Then $\mu=Comp(\Sigma_1,\Pi_1)$ on pseudotrees. By
 Lemma \ref{lemma:pseudo}, every finite graph belonging to $SCC1$ is
 bisimilar to a finite pseudotree.   So, by our
 hypothesis and by invariance of the $\mu$-Calculus under
 bisimulation, we would have that $\mu=Comp(\Sigma_1,\Pi_1)$
   on $SCC1$. But this is in contrast with Corollary \ref{cor:coll}.

 \rightline{Q.E.D.}

\medskip

By the previous corollary, we have a lower bound on the $\mu$-calculus hierarchy on the class $TW1$, hence also on the larger classes $TWk$ for every $k>1$.
 It would be interesting to come up with an {\em upper} bound on $TWk$ as well,
 and more generally, to investigate the expressiveness of $\mu$-Calculus on classes given by other, algorithmically interesting graph-theoretic measures (e.g. cliquewidth, DAG-width, etc.) This will be the subject of future papers.

\section*{Acknowledgments}
The work has been partially supported by 
the PRIN project \emph{Innovative and multi-disciplinary
approaches for constraint and preference reasoning}
and the GNCS project \emph{Logics, automata, and games
for the formal verification of complex systems}.

\bibliographystyle{eptcs}

\begin{thebibliography}{00}

\bibitem{AF} L. Alberucci and A. Facchini,
The Modal mu-Calculus Hierarchy over Restricted Classes of Transition Systems,
Journal of Symbolic Logic, (4) 74 (2009), 1367--1400.

\bibitem{AF09} L. Alberucci and A. Facchini,
On Modal $\mu$-Calculus and G\"odel-L\"ob Logic, Studia Logica 91 (2009), 145--169.

\bibitem{A99} A. Arnold, The mu-Calculus Alternation-Depth Hierarchy is Strict on Binary Trees. ITA (33) 4/5 (1999), 329--340.

\bibitem{B96} J. C. Bradfield, The Modal mu-Calculus Alternation Hierarchy is Strict, Proceedings of CONCUR 1996, 233--246.

\bibitem{MC} E. M. Clarke, O. Grumberg and D. A. Peled, Model Checking,
MIT Press, 1999.

\bibitem{Cou90} B. Courcelle, Graph rewriting: An algebraic and logic approach, In: J. van
Leeuwen, editor, Handbook of Theoretical Computer Science: Volume B: Formal
Models and Semantics, Elsevier, Amsterdam, 1990, 193--242.

\bibitem{DL10} G. D'Agostino and G. Lenzi, On the $\mu$-Calculus over Transitive and Finite
Transitive Frames, submitted.

\bibitem{DO09} A. Dawar and M. Otto,
Modal Characterisation Theorems over Special Classes of Frames,
Annals of Pure and Applied Logic 161 (2009), 1--42.

\bibitem{EJ} E. A. Emerson and C. S. Jutla, Tree Automata, Mu-Calculus and Determinacy, IEEE Proc. Foundations of Computer Science (1991), 368--377.

\bibitem{EL86} E. A. Emerson and C. L. Lei. Efficient Model Checking in Fragments of the
Propositional $\mu$-Calculus. In: Symposium on Logic in Computer Science, pages
267�278. IEEE Computer Society Press, June 1986.

\bibitem{FG} J. Flum, M. Grohe, Parameterized Complexity Theory, Springer, 2006.

\bibitem{H76} R. Halin, S-Functions for Graphs, J. Geometry 8 (1976), 171--186.

\bibitem{JW96} D. Janin, I. Walukiewicz, On the Expressive Completeness of the Propositional mu-Calculus with Respect to Monadic Second Order Logic, CONCUR 1996,  263--277.

\bibitem{JRST} T. Johnson, N. Robertson, P. D. Seymour and R. Thomas, Directed Tree-Width, J. Combin. Theory Ser. B 82 (2001), 138--155.

\bibitem{J98} M. Jurdzi{\'n}ski, Deciding the Winner in Parity Games Is in
                  $UP\cap~co-UP$,
  Information Processing Letters (68) 3 (1998), 119--124.

\bibitem{JPZ} M. Jurdzi{\'n}ski, M. Paterson, U. Zwick, A Deterministic Subexponential Algorithm for Solving Parity Games, SIAM J. Comput. (4) 38 (2008), 1519--1532.

\bibitem{K83} D. Kozen, Results on the Propositional $\mu$-Calculus. Theor. Comput. Sci. 27 (1983), 333--354.

\bibitem{VK}  O. Kupferman, M. Y. Vardi: $\Pi_2\cap\Sigma_2\equiv AFMC$. Proceedings of ICALP 2003, 697--713.

\bibitem{LBC+94} D. E. Long, A. Browne, E. M. Clarke, S. Jha, and W. R. Marrero, An
Improved Algorithm for the Evaluation of Fixpoint Expressions, In CAV '94,
volume 818 of LNCS, Springer-Verlag, 1994, 338�-350.

\bibitem{M75} D. Martin, Borel Determinacy, Annals of Mathematics. Second series (2) 102 (1975), 363--371.

\bibitem{O03} J. Obrd\v{z}{\'a}lek, Fast Mu-Calculus Model Checking when Tree-Width Is Bounded, Proceedings of CAV 2003, 80--92.

\bibitem{R69} M. Rabin, Decidability of Second-Order Theories and Automata on Infinite Trees, Transactions of the American Mathematical Society 141 (1969), 1--35.

\bibitem{RS} N. Robertson and P. Seymour,   Graph Minors III: Planar Tree-Width, Journal of Combinatorial Theory, Series B, vol. 36 (1984), 49--64.

\bibitem{RS86}  N. Robertson and P. D. Seymour, Graph Minors. V. Excluding a Planar Graph, J. Combin. Theory Ser. B 41 (1986), 92--114.

\bibitem{Smo85} C. Smory{\'n}ski, Self-reference and Modal Logic, Springer, 1985.

\bibitem{VB76} J. van Benthem, Modal Correspondence Theory, Ph.D. Thesis, Mathematisch Instituut \& Instituut voor Grondslagenonderzoek,
University of Amsterdam, 1976.

\bibitem{VB06} J. Van Benthem, Modal Frame Correspondences and Fixed Points, Studia Logica 83 (2006), 133--155.

\bibitem{V05} A. Visser, L\"ob's Logic Meets the $\mu$-Calculus, in: A. Middeldorp, V. van Oostrom, F. van Raamsdonk and R. de Vrijer (eds.), Processes, Terms and Cycles, Steps on the Road to Infinity, Essays Dedicated to Jan Willem Klop on the Occasion of his 60th Birthday, Springer, 2005, 14--25.

\end{thebibliography}

\end{document}